\documentclass[12pt,fleqn]{article}

\usepackage{amsmath,amssymb}

\def\tlabel#1{\label{#1}}
\def\tbib#1#2#3{{#1}~(#2)~#3}
\def\tref#1{(\ref{#1})}
\def\unt#1{\,{\rm #1}}

\begin{document}
\title{
{\sf Two Topics in QCD Sum-Rules: Bounds on Light Quark and Scalar Gluonium Masses}
}
\author{
 T.G.~Steele\thanks{email: Tom.Steele@usask.ca} ~and D. Harnett\\
  \textsl{Department of Physics and Engineering Physics}\\
  \textsl{University of Saskatchewan}\\
  \textsl{Saskatoon, Saskatchewan, S7N 5E2}\\
  \textsl{Canada}
}
\maketitle

\begin{abstract}
QCD Laplace sum-rules are briefly reviewed and two sum-rule applications are presented.  
For scalar gluonium, upper bounds on the lightest 
state are obtained from ratios of Laplace sum-rules, and the 
role of instantons and the low-energy theorem on these bounds is 
investigated.  For the light-quark masses, fundamental inequalities 
for Laplace sum-rules are developed, which lead to a lower bound on the average of the 
up and down quark masses.
\end{abstract}

\section{Introduction}\label{IntroSec}
QCD sum-rules, particularly in the form of Laplace sum-rules \cite{SVZ}, 
are a well-established technique for relating QCD predictions to hadronic physics.   
In this approach, the non-perturbative aspects of the QCD vacuum are parametrized indirectly through the
QCD condensates, or directly by instanton contributions.  

In this paper the QCD Laplace sum-rule approach is briefly reviewed in Section \ref{SumRuleSec}, 
and two applications are presented
with the common theme of obtaining mass bounds.  In Section \ref{GlueSec}, QCD sum-rules mass bounds on the lightest scalar glueball state will be obtained.  Instanton effects 
in the scalar gluonium sum-rules are seen to be important, and help reconcile a number of 
conflicting sum-rule results in this channel.  Finally, Section \ref{MassBoundsSec} develops 
fundamental inequalities for QCD sum-rules which are then applied to obtain mass bounds for the
non-strange light quarks.

\section{QCD Laplace Sum-Rules}\label{SumRuleSec}
QCD sum-rules are based upon correlation functions of gauge invariant and renormalization group (RG) invariant composite operators
\begin{equation}
\Pi_{_\Gamma}\left(Q^2\right)=i\int \mathrm{d}^4x\, e^{iq\cdot x}\left\langle O \vert T\left[ J_{_\Gamma}(x) J_{_\Gamma}(0)\right] \vert O \right\rangle\quad ,
\tlabel{GenCorrFn}
\end{equation}
with $J_{_\Gamma}(x)$  a composite operator with desired quantum numbers  $\Gamma$.  These composite
operators serve as interpolating fields between  hadronic states and the vacuum, appropriate to the quantum numbers of the current.

The  correlation function \tref{GenCorrFn} satisfies a dispersion relation  with several subtractions appropriate to the 
large-energy asymptotic behaviour\footnote{The given form is appropriate to scalar gluonium.} of the correlation function  
\begin{equation}
\Pi_\Gamma\left(Q^2\right)=\Pi_\Gamma(0)+Q^2\Pi'_\Gamma(0)+\frac{1}{2}Q^4\Pi''_\Gamma(0)
-Q^6\frac{1}{\pi}\int\limits_{t_0}^\infty
\,\frac{ \rho_{_\Gamma}(t)}{t^3\left(t+Q^2\right)}\mathrm{d}t \quad ,
\tlabel{GenDispRel}
\end{equation}
where $\rho_{_\Gamma}(t)$ is the hadronic spectral function with physical threshold $t_0$ appropriate to the quantum numbers of the currents.  The dispersion relation serves to relate the (Euclidean momenta)
QCD prediction $\Pi_\Gamma\left(Q^2\right)$ to hadronic spectral function.  The subtraction constants 
$\{\Pi_\Gamma(0), \Pi'_\Gamma(0),\Pi''_\Gamma(0)\}$
are usually unknown, but in certain cases they are determined by low-energy theorems \cite{NSVZ}.

The dispersion relations are not suitable for applications to light hadronic channels since they 
receive substantial contributions from excited states and the QCD continuum in the integral of 
$\rho_{_\Gamma}(t)$, obscuring the lightest states.   These difficulties were resolved by applying the Borel transform operator \cite{SVZ}
 to the dispersion relation
\begin{equation}
\hat B\equiv 
\lim_{\stackrel{N,~Q^2\rightarrow \infty}{N/Q^2\equiv \tau}}
\frac{\left(-Q^2\right)^N}{\Gamma(N)}\left(\frac{\mathrm{d}}{\mathrm{d}Q^2}\right)^N\quad .
\tlabel{BorelOp}
\end{equation}
This operator  has the following properties useful to construction of Laplace sum-rules:
\begin{gather}
\hat B\left[a_0+a_1Q^2+\ldots a_m Q^{2m}\right]=0\quad,\quad (m~{\rm finite})
\tlabel{BorelPoly}\\
 \hat B \left[ \frac{Q^{2n}}{t+Q^2}\right]=\tau \left(-1\right)^nt^n\mathrm{e}^{-t\tau}  \quad,\quad n=0,~1,~2,\ldots ~ 
(n~{\rm finite})\quad .
\tlabel{BorelExp}
\end{gather}
In general, $\hat B$  is related to the inverse Laplace transform  \cite{borel_ref}
\begin{gather}
f\left(Q^2\right)=\int\limits_0^\infty F(\tau) \mathrm{e}^{-Q^2\tau}\, \mathrm{d}\tau\equiv{\cal L}\left[ F(\tau)\right]
~ \Longrightarrow~\frac{1}{\tau}\hat B\left[ f\left(Q^2\right)\right]
=F(\tau)={\cal L}^{-1}
\left[ f\left(Q^2\right)\right]
\tlabel{BorelLaplace}\\
{\cal L}^{-1}
\left[ f\left(Q^2\right)\right]=\frac{1}{2\pi i}\int\limits_{b-i\infty}^{b+i\infty}
f\left(Q^2\right) \mathrm{e}^{Q^2\tau}\,\mathrm{d}Q^2\quad ,
\tlabel{InvLapDef}
\end{gather}
where the real parameter $b$ in the definition \tref{InvLapDef} of the inverse Laplace 
transform must be chosen so that $f\left(Q^2\right)$ is analytic to the right of the 
contour of integration in the complex plane.

Families of Laplace sum-rules are now obtained by applying $\hat B$ to \tref{GenDispRel} weighted by integer powers of $Q^2$, which will involve the theoretically-determined quantity
\begin{eqnarray}
{\cal L}^{(\Gamma)}_k(\tau)\equiv\frac{1}{\tau}\hat B\left[\left(-1\right)^k Q^{2k}\Pi_{_\Gamma}\left(Q^2\right)\right]
={\cal L}^{-1}\left[
\left(-1\right)^k Q^{2k}\Pi_\Gamma\left(Q^2\right)
\right]\quad .
\tlabel{laplace}
\end{eqnarray}
For $k\ge 0$, this results in the following Laplace sum-rules relating QCD to hadronic physics\footnote{The case $k< 0$ is 
relevant to scalar gluonium, and will be discussed in Section \protect\ref{GlueSec}.}
\begin{equation}
{\cal L}^{(\Gamma)}_{k}(\tau)=\frac{1}{\pi}\int\limits_{t_0}^\infty
t^k \mathrm{e}^{-t\tau}\rho_{_\Gamma}(t)\,\mathrm{d}t \quad,\quad k\ge 0\quad .
\tlabel{GenLap}
\end{equation}
 Compared with the dispersion relation \tref{GenDispRel},
the exponential weight in \tref{GenLap} suppresses the (large-energy) QCD continuum and excited states
in the Laplace sum-rule, emphasizing the lightest resonances in a particular channel.  The quantity $\tau$ can truly be thought of as a scale which probes the spectral function, since in addition to controlling the region of
integration on the hadronic (right-hand) side of  \tref{GenLap}, RG improvement of the QCD (left-hand) side of \tref{GenLap} implies that the renormalization scale is $\nu^2=1/\tau$ \cite{bounds}.

Analysis of the QCD Laplace sum-rules is achieved through a resonance(s) plus continuum model, where 
hadronic physics is locally dual to QCD for energies above the continuum threshold $t=s_0$
\begin{equation}
\rho_{_\Gamma}(t)=\rho_{_\Gamma}^{had}(t)+\theta\left(t-s_0\right){\rm Im}\Pi_{_\Gamma}^{QCD}(t)
\quad .
\tlabel{ResPlusCont}
\end{equation}
The QCD continuum contribution denoted by
\begin{equation}
c^{(\Gamma)}_k\left(\tau,s_0\right)=\frac{1}{\pi}\int\limits_{s_0}^\infty
\,t^k \mathrm{e}^{-t\tau}{\rm Im}\Pi_{_\Gamma}^{QCD}(t)\,\mathrm{d}t\quad ,
\tlabel{Continuum}
\end{equation}
 is combined with ${\cal L}^{(s)}_k(\tau)$ since both are theoretically determined
\begin{equation}
R^{(\Gamma)}_k\left(\tau,s_0\right)\equiv {\cal L}^{(\Gamma)}_k\left(\tau\right)-
c^{(\Gamma)}_k\left(\tau,s_0\right)\quad ,
\tlabel{FullLap}
\end{equation}
resulting in  the final form of the Laplace sum-rules relating QCD to hadronic physics phenomenology
\begin{equation}
R^{(\Gamma)}_{k}\left(\tau,s_0\right)=\frac{1}{\pi}\int\limits_{t_0}^{\infty}
\,t^k \mathrm{e}^{-t\tau}\rho^{had}_{_\Gamma}(t)\,\mathrm{d}t\quad ,\quad k\ge 0\quad .
\tlabel{FullRule}
\end{equation}
After a hadronic model for $\rho^{had}_{_\Gamma}(t)$ is established, its resonance parameters and the 
continuum threshold $s_0$ are determined by fitting the two sides of \tref{FullRule} over a range of 
the Laplace scale $\tau$ probing the theory and phenomenology.

\section{Mass Bounds on the Lightest Scalar Gluonium State}\label{GlueSec}
Scalar gluonium is studied through the  current
\begin{gather}
J_s(x)=-\frac{\pi^2}{\alpha\beta_0}\beta\left(\alpha \right)G^a_{\mu\nu}(x)G^a_{\mu\nu}(x)
\tlabel{GlueCurrent}\\
\beta\left(\alpha\right) =\nu^2\frac{\mathrm{d}}{\mathrm{d}\nu^2}\left(\frac{\alpha(\nu)}{\pi}\right)=
-\beta_0\left(\frac{\alpha}{\pi}\right)^2-\beta_1\left(\frac{\alpha}{\pi}\right)^3+\ldots
\tlabel{beta}\\
\beta_0=\frac{11}{4}-\frac{1}{6} n_f\quad ,\quad
\beta_1=\frac{51}{8}-\frac{19}{24}n_f
\quad .
\tlabel{beta_coeffs}
\end{gather}
which is RG invariant for massless quarks \cite{RG}, 
and where the current normalization has been chosen 
so that to lowest order in $\alpha$
\begin{equation}
J_s(x)=\alpha G^a_{\mu\nu}(x)G^a_{\mu\nu}(x)\left[1+\frac{\beta_1}{\beta_0}\frac{\alpha}{\pi}
+{\cal O}\left(\alpha^2\right)\right]
\equiv \alpha G^2(x)\left[1+\frac{\beta_1}{\beta_0}\frac{\alpha}{\pi}+{\cal O}\left(\alpha^2\right)\right]
\quad .
\tlabel{CurrentExp}
\end{equation}
The correlation function of $J_s$ has the interesting property that
in the chiral limit of $n_f$ quarks, it satisfies the following  low-energy theorem (LET) 
\cite{NSVZ}
\begin{equation}
\Pi_s(0)=\lim_{Q\rightarrow 0} \Pi_s\left(Q^2\right)=\frac{8\pi}{\beta_0}\left\langle J_s\right\rangle\quad .
\tlabel{let}
\end{equation}
This correlation function satisfies a dispersion relation identical in form to \tref{GenDispRel}
($\Gamma$ is identified with $s$), implying that the first subtraction constant $\Pi_s(0)$ is determined by the LET.  This property allows extension of the sum-rule \tref{GenLap} to include $k=-1$ at the expense of introducing the LET term:\footnote{For $k\le -2$ the sum-rules would have dependence on subtraction constants such as $\Pi'_\Gamma(0)$ 
{\em not} determined by the LET.}
\begin{gather}
{\cal L}^{(s)}_{-1}(\tau)={-\Pi_s(0)}+\frac{1}{\pi}\int\limits_{t_0}^\infty
\,\frac{1}{t} \mathrm{e}^{-t\tau}\rho_s(t)\,\mathrm{d}t
\tlabel{GlueM1Lap}\\
{\cal L}^{(s)}_{k}(\tau)=\frac{1}{\pi}\int\limits_{t_0}^\infty
t^k \mathrm{e}^{-t\tau}\rho_s(t)\,\mathrm{d}t\quad ,\quad k\ge 0\quad .
\tlabel{GluekLap}
\end{gather}

The only appearance of the $\Pi_s(0)$ term is in the $k=-1$ sum-rule, and as first noted in
\cite{NSVZ_glue}, this LET term comprises a significant  contribution in this sum-rule.
From the significance of this ($\tau$) scale-independent term one can ascertain the  
important qualitative role of the LET in  sum-rule phenomenology.  
To see this role, we first model the hadronic contributions 
 $\rho_s^{had}(t)$ using the narrow resonance approximation
\begin{equation}
\frac{1}{\pi}\rho^{had}(t)=\sum_r F_r^2m_r^2\delta\left(t-m_r^2\right)\quad ,
\tlabel{narrow}
\end{equation}
where the  sum over $r$ represents a sum over  resonances of mass $m_r$.  The quantity
 $F_r$ is the coupling strength of the resonance to the vacuum through the gluonic current $J_s(0)$, so
the sum-rule for scalar gluonic currents probes  scalar gluonium states.  
In the narrow-width approximation the Laplace sum-rules 
become
\begin{gather}
R^{(s)}_{-1}\left(\tau,s_0\right)+\Pi_s(0)=
\sum_r F_r^2 \mathrm{e}^{-m_r^2\tau}
\tlabel{lap_m1_res}\\
R^{(s)}_{k}\left(\tau,s_0\right)=
\sum_r F_r^2m_r^{2k+2} \mathrm{e}^{-m_r^2\tau}\quad ,\quad k\ge 0\quad .
\tlabel{LapkRes}
\end{gather}
Thus if the (constant) LET term is a significant contribution on the theoretical side of \tref{lap_m1_res}, then
the left-hand side of \tref{lap_m1_res} will exhibit reduced $\tau$ dependence relative to other theoretical 
contributions.  To reproduce this diminished $\tau$ dependence, the phenomenological 
({\it i.e.} right-hand)
side must 
contain a  light resonance with a coupling larger than or comparable to the heavier resonances.  
By contrast, the absence of the $\Pi_s(0)$ (constant) term in $k>-1$ sum-rules leads to stronger $\tau$ dependence
which is balanced on the phenomenological side  by suppression 
of the lightest resonances 
via the additional powers of $m_r^2$  occurring in \tref{LapkRes}.  Thus if $\Pi_s(0)$ is found to dominate 
$R^{(s)}_{-1}\left(\tau,s_0\right)$, then one would expect qualitatively different results from analysis
of the $k=-1$ and $k>-1$ sum-rules.  

Such distinct conclusions drawn from different sum-rules can be legitimate.  
In the pseudoscalar quark sector, the lowest sum-rule is dominated by the pion, and the low  mass of the pion 
is evident from the  minimal $\tau$ dependence
in the lowest sum-rule. By contrast,  the first subsequent sum-rule has an important contribution from the 
$\Pi(1300)$ \cite{pi_1300},  as the pion contribution is suppressed by its low mass, resulting in the 
significant $\tau$ dependence of the next-to-lowest sum-rule.

In the absence of instantons \cite{instanton}, 
explicit sum-rule analyses of scalar gluonium \cite{NSVZ_glue,glue} uphold the above 
generalization---those which include the
$k=-1$ sum-rule find a light (less than or on the order of the $\rho$ mass) gluonium state, 
and those which omit the $k=-1$  sum-rule find a state with a mass greater than $1\unt{GeV}$. 
The prediction of a light gluonium state  would have interesting 
phenomenological consequences as a state which could be identified with the $f_0(400-1200)$ (or $\sigma$) 
meson \cite{ochs}.  

Work by Shuryak \cite{shuryak} in the instanton liquid model \cite{ins_liquid} has indicated how
a large-energy asymptotic expression for the instanton contribution to the $k=-1$ sum-rule
may serve to compensate for that sum-rule's LET component and bring the predicted scalar gluonium mass in
line with subsequent lattice estimates ($\sim 1.6\unt{GeV}$ \cite{lattice}).  Recent work by Forkel \cite{forkel} has addressed
in detail instanton effects on scalar gluonium mass predictions from higher-weight ($k\ge 0$) sum-rules and has also corroborated
lattice estimates.  The overall consistency of the LET-sensitive $k=-1$ sum-rule and the LET-insensitive $k\ge 0$ sum-rules has been addressed in \cite{ourglue}, and will be reviewed in the remainder of this section.

The field-theoretical (QCD) calculation of $\Pi\left(Q^2\right)$ consists of perturbative (logarithmic) corrections 
known to three-loop order ($\overline{{\rm MS}}$ scheme) in the chiral limit of $n_f=3$ massless quarks 
\cite{glue_pt}, 
QCD vacuum effects of infinite correlation length 
parameterized by  the power-law contributions from the QCD vacuum condensates 
\cite{NSVZ_glue,condensate2}, 
\footnote{The calculation of one-loop contributions proportional to $\left\langle J_s\right\rangle$ 
in \protect\cite{condensate2} have been extended 
non-trivially
to $n_f=3$ from $n_f=0$, and the operator basis
has been changed from $\left\langle \alpha G^2\right\rangle$ to $\left\langle J_s\right\rangle$. 
}
and QCD vacuum effects of finite correlation length devolving from instantons \cite{instanton_corr}
\begin{equation}
\Pi_s\left(Q^2\right)=
\Pi_s^{pert}\left(Q^2\right)+\Pi_s^{cond}\left(Q^2\right)+\Pi_s^{inst}\left(Q^2\right)\quad .
\tlabel{QCD_corr_fn}
\end{equation}
 The perturbative contribution (ignoring divergent terms proportional to $Q^4$) to \tref{QCD_corr_fn} is 
\begin{gather}
\Pi_s^{pert}\left(Q^2\right)=Q^4\log\left(\frac{Q^2}{\nu^2}\right)\left[
a_0+a_1\log\left(\frac{Q^2}{\nu^2}\right)+a_2\log^2\left(\frac{Q^2}{\nu^2}\right)
\right]
\tlabel{pi_pert}\\
a_0=-2\left(\frac{\alpha}{\pi}\right)^2\left[1+\frac{659}{36}\frac{\alpha}{\pi}+
247.480\left( \frac{\alpha}{\pi}\right)^2\right]
~ , ~  
a_1=2\left(\frac{\alpha}{\pi}\right)^3\left[ \frac{9}{4}+65.781\frac{\alpha}{\pi}\right]
\nonumber\\
a_2=-10.1250\left(\frac{\alpha}{\pi}\right)^4\quad ,
\nonumber
\end{gather}
the condensate contributions to \tref{QCD_corr_fn} are  
\begin{gather}
\Pi_s^{cond}\left(Q^2\right)=
\left[ b_0+b_1\log\left(\frac{Q^2}{\nu^2}\right)\right]\left\langle  J_s\right\rangle
+c_0\frac{1}{Q^2}\left\langle {\cal O}_6\right\rangle+d_0\frac{1}{Q^4}\left\langle {\cal O}_8\right\rangle
\tlabel{pi_cond}\\
b_0=4\pi\frac{\alpha}{\pi}\left[ 1+ \frac{175}{36}\frac{\alpha}{\pi}\right]
~ ,~ b_1=-9\pi\left(\frac{\alpha}{\pi}\right)^2
~ ,~
c_0=8\pi^2\left(\frac{\alpha}{\pi}\right)^2~ ,~ d_0=8\pi^2\frac{\alpha}{\pi}
\nonumber
\\
\left\langle {\cal O}_6\right\rangle=
\left\langle g f_{abc}G^a_{\mu\nu}G^b_{\nu\rho}G^c_{\rho\mu}\right\rangle
~ ,~
\left\langle {\cal O}_8\right\rangle=14\left\langle\left(\alpha f_{abc}G^a_{\mu\rho}G^b_{\nu\rho}\right)^2\right\rangle
-\left\langle\left(\alpha f_{abc}G^a_{\mu\nu}G^b_{\rho\lambda}\right)^2\right\rangle
\nonumber
\end{gather}
 and finally the instanton contribution to \tref{QCD_corr_fn} is  
\begin{equation}
\Pi_s^{inst}\left(Q^2\right)=
32\pi^2Q^4\int \rho^4 \left[K_2\left(\rho\sqrt{Q^2}\right)\right]^2  \mathrm{d}n(\rho)\quad ,
\tlabel{pi_inst}
\end{equation}
where $K_2(x)$ represents a modified Bessel function \cite{abramowitz}.
The instanton contributions 
represent a calculation with non-interacting instantons 
of size $\rho$, with subsequent integration over the instanton density distribution $n(\rho)$. 

The strong coupling constant $\alpha$ in \tref{pi_pert} is understood to be the running coupling at the renormalization 
scale $1/\sqrt{\tau}$, and its 
 three-loop, $n_f=3$, $\overline{\rm MS}$ running form is :  
\begin{gather}
\frac{\alpha_s(\nu)}{\pi}= \frac{1}{\beta_0 L}-\frac{\bar\beta_1\log L}{\left(\beta_0L\right)^2}+
\frac{1}{\left(\beta_0 L\right)^3}\left[
\bar\beta_1^2\left(\log^2 L-\log L -1\right) +\bar\beta_2\right]
\label{alpha_hl}\\
L=\log\left(\frac{\nu^2}{\Lambda^2}\right)~,~\bar\beta_i=\frac{\beta_i}{\beta_0}
~,
~
\beta_0=\frac{9}{4}~,~\beta_1=4~,~\beta_2=\frac{3863}{384}
\end{gather}
with 
 $\Lambda_{\overline{MS}}\approx 300\,{\rm MeV}$ for three active flavours,
consistent with current estimates of $\alpha_s(M_\tau)$ \cite{aleph}. 
The dimension-six and dimension-eight gluon condensates are referenced to the gluon condensate $\langle \alpha G^2\rangle$ through  vacuum saturation \cite{NSVZ_glue,vacuum_saturation}
\begin{equation}
\left\langle {\cal O}_8\right\rangle=14\left\langle\left(\alpha f_{abc}G^a_{\mu\rho}G^b_{\nu\rho}\right)^2\right\rangle
-\left\langle\left(\alpha f_{abc}G^a_{\mu\nu}G^b_{\rho\lambda}\right)^2\right\rangle=
\frac{9}{16}\left(\left\langle \alpha G^2\right\rangle\right)^2\quad ,
\label{vac_sat}
\end{equation}
and  instanton estimates \cite{SVZ,NSVZ} 
\begin{equation}
\left\langle {\cal O}_6\right\rangle=
\left\langle g f_{abc}G^a_{\mu\nu}G^b_{\nu\rho}G^c_{\rho\mu}\right\rangle
=\left(0.27\unt{GeV^2}\right)\left\langle \alpha G^2\right\rangle\quad .
\end{equation}

The Laplace sum-rules can also be partitioned into perturbative, condensate and instanton contributions 
 \begin{equation}
{\cal L}^{(s)}_k(\tau)=
{\cal L}_k^{pert}(\tau)+{\cal L}_k^{cond}(\tau)+{\cal L}_k^{inst}(\tau) \quad.
\tlabel{laplace_ports}
\end{equation}
Calculation of the  instanton contributions to these  Laplace sum-rules demands particular care, and requires 
the use of the inverse Laplace transform in the complex $Q^2$ plane \tref{laplace} to calculate the Borel transform.
The complete details of this calculation are given in \cite{ourglue}, but the essential aspect of the result is 
that the instanton contributions distinguish between $k=-1$ and $k>-1$ sum-rules, similar to the unique role played by the LET term in the $k=-1$ sum-rule \cite{ourglue}.
\begin{gather}
{\cal L}_{-1}^{inst}\left(\tau\right)=
{-128\pi^2\int \mathrm{d}n(\rho)}
-16\pi^3\int \mathrm{d}n(\rho) \rho^4\int\limits_0^\infty tJ_2\left(\rho\sqrt{t}\right)
Y_2\left(\rho\sqrt{t}\right)\mathrm{e}^{-t\tau} \, \mathrm{d}t 
\tlabel{LapInstM1}       \\
{\cal L}_k^{inst}\left(\tau\right)=
-16\pi^3\int \mathrm{d}n(\rho) \rho^4\int\limits_0^\infty t^{k+2}J_2\left(\rho\sqrt{t}\right)
Y_2\left(\rho\sqrt{t}\right)\mathrm{e}^{-t\tau} \, \mathrm{d}t \quad ,\quad k>-1 
\tlabel{LapInstk}
\end{gather}
These results are   interpreted as a natural partitioning of  instanton  contributions into an
instanton continuum portion devolving from
\begin{equation}
\frac{1}{\pi}{\rm Im}\Pi_s^{inst}(t)=-16\pi^3\int \mathrm{d}n(\rho)\,\rho^4t^2
J_2\left(\rho\sqrt{t}\right) Y_2\left(\rho\sqrt{t}\right) \quad ,
\tlabel{InsCont}
\end{equation}
and a contribution which, like the LET, appears only in the $k=-1$ sum-rule
\begin{equation}
-128\pi^2 \int \mathrm{d}n(\rho)\quad .
\tlabel{InstLET}
\end{equation}
 
The qualitative role of instanton effects can be investigated in the instanton liquid model \cite{ins_liquid}
\begin{equation}
\mathrm{d}n(\rho)=n_c\delta\left(\rho-\rho_c\right)\mathrm{d}\rho
~,~ n_c=8\times 10^{-4}\unt{GeV^4}~ ,~
\rho_c=\frac{1}{600\unt{MeV}}
\tlabel{InsLiqModel}
\end{equation}
Thus in the $k=-1$ sum-rule \tref{GlueM1Lap} a cancellation occurs between LET and the LET-like instanton contribution
\tref{InstLET}
\begin{gather}
\frac{-\Pi_s(0)+128\pi^2 \int \mathrm{d}n(\rho)}{\Pi_s(0)}\approx -1+\frac{128\pi^2n_c}{\Pi_s(0)}\le
0.29\pm 0.16
\tlabel{Cancel}
\end{gather}
where we have used 
\begin{equation}
\Pi_s(0)=\frac{32\pi}{9}\left\langle J_s\right\rangle
\ge \frac{32\pi}{9}\left\langle \alpha G^2\right\rangle
\tlabel{LET_value}
\end{equation}
in conjunction with a recent determination of the
of the gluon condensate $\langle \alpha G^2\rangle$ \cite{narison2}\footnote{
It should be noted that there is some discrepancy between 
\protect\cite{narison2} 
 and the smaller value $\langle\alpha G^2\rangle=\left( 0.047\pm 0.014\right)\,{\rm GeV^4}$ found in \protect\cite{dimsix2}. 
}
\begin{equation}
\langle \alpha G^2\rangle=\left(0.07\pm 0.01\right)\,{\rm GeV^4}\quad .
\tlabel{aGG}
\end{equation}
Thus we see that instantons could suppress the role of the LET by 55--85\%, reducing the discrepancy between
gluonium mass determinations in the $k=-1$ and $k\ge 0$   sum-rule analyses  \cite{ourglue}.

This qualitative conclusion is borne out in more detailed analysis.  
Within the instanton liquid model, the instanton contribution \tref{LapInstM1} becomes \cite{ourglue}
\begin{equation}
{\cal L}_{-1}^{inst}(\tau)=-64\pi^2n_c
\left[a^2e^{-a}\left[a+1\right]K_0(a)\!+\!ae^{-a}\left[a^2+2a+2\right]K_1(a)\right]
~,~a=\frac{\rho_c^2}{2\tau}~.
\tlabel{FullM1Inst}
\end{equation}
The large-energy ($a\gg 1$) limit  of \tref{FullM1Inst} agrees with the asymptotic form 
\begin{equation}
{\cal L}_{-1}^{inst}(\tau)
= -16\pi^{\frac{5}{2}}n_c  \mathrm{e}^{-\frac{\rho_c^2}{\tau}}\rho_c^5\tau^{-\frac{5}{2}}
\left[1+{\cal O}\left(\frac{\tau}{\rho_c^2}\right)\right]~,~\tau\ll\rho_c^2~.
\tlabel{ShuryakForm}
\end{equation}
given in \cite{shuryak}.  However, numerical comparison with \tref{FullM1Inst} reveals that 
the higher-order terms omitted in \tref{ShuryakForm} are significant in the region 
$a<2$ ($\tau>0.54\,{\rm GeV^2}$), an energy region important to a sum-rule analysis, so it is necessary to use the full expression \tref{FullM1Inst} for an accurate analysis of instanton effects.  
Conversely, the small $a$ limit of \tref{FullM1Inst}
\begin{equation}
{\cal L}_{-1}^{inst}(\tau) \rightarrow
-128\pi^2n_c
+8\pi^2n_c \left[
\frac{\rho_c^4}{\tau^2}+{\cal O}\left( \frac{\rho_c^6}{\tau^3}\right)
\right]\quad ,\quad \tau\gg\rho_c^2
\tlabel{SmallForm}
\end{equation}
upholds the interpretation of the partitioning of the instanton contribution into an LET-like contribution
[the first term on the right-hand side of \tref{SmallForm}], while remaining a good numerical approximation to the 
full expression \tref{FullM1Inst} in the region  $\tau\gtrsim 0.85\,{\rm GeV^{-2}}$.

Ratios of Laplace sum-rules provide a bound on the mass $m$ of the lightest state contributing to the spectral function.  For scalar gluonium, some of these possible ratios are\footnote{Derivation of these bounds, along with the complete sum-rule expressions needed to analyze these ratios are given in ref.\ \cite{ourglue}.}
\begin{gather}
\frac{{\cal L}^{(s)}_1\left(\tau\right)}{{\cal L}^{(s)}_0\left(\tau\right)}
\ge {m^2}
\tlabel{NonLetBound}\\[10pt]
\frac{{\cal L}^{(s)}_0\left(\tau\right)}{{\cal L}^{(s)}_{-1}\left(\tau\right)+\Pi_s(0)}
\ge {m^2} \quad .
\tlabel{LetBound}
\end{gather}
In the absence of instantons the mass bounds obtained from the ratio \tref{LetBound} containing the LET are much 
lower than the bound obtained from the ratio \tref{NonLetBound} where the LET term is absent.  However, the 
full form of the instanton contributions {\em raises} the mass bound from the LET-dependent ratio 
\tref{LetBound}, and 
{\em lowers} the mass bound from the LET-independent ratio  \tref{NonLetBound}, leading
to consistent  predictions from the entire family of sum-rules \cite{ourglue}.  With the inclusion of
instanton effects, the final mass bounds from the sum-rule ratios are approximately \cite{ourglue}
\begin{equation}
m\lesssim 1.25\,{\rm GeV}\quad ,
\end{equation}
a result consistent with a recent  Gaussian sum-rule analysis of scalar gluonium \cite{newglue}. 
These results suggest that there exists a (possibly small) gluonic component of the PDG-listed states
$f_0(400\mbox{--}1200)$, $f_0(980)$, or $f_0(1370)$.

\section{Bounds on the Light Quark Masses}\label{MassBoundsSec}
The light quark masses are  fundamental parameters of QCD, and 
determination of their values is of importance for high-precision 
QCD phenomenology and lattice simulations involving dynamical quarks.
In this paper the development of H\"older inequalities for QCD Laplace sum-rules \cite{sr_holder} 
is briefly reviewed.  These  techniques are then used to obtain bounds on the 
non-strange (current) quark masses $m_n=\left(m_u+m_d\right)/2$  evaluated at $2\,{\rm GeV}$ in the $\overline{\rm MS}$ scheme, updating and
extending the H\"older inequality results of ref.\ \cite{holder_bounds}.

Although it is possible to obtain quark mass ratios in various contexts \cite{mass_ratios}, the only methods
which have been able to determine the {\em absolute} non-strange quark mass scales are the 
lattice (see \cite{lattice_mass} for recent results with two dynamical flavours)  
and QCD sum-rules 
\cite{bounds,pi_1300,holder_bounds,BNRY,other_bounds,sumrule_mass,3pi}.\footnote{An overview of selected lattice and sum-rule results for both non-strange and strange masses can be found  in \protect\cite{gupta}.}

In sum-rule and lattice  approaches, the pseudoscalar or scalar channels are used since they have the 
strongest dependence on the quark masses.  
This is exemplified by the correlation function 
 $\Pi_5\left(Q^2\right)$  of 
renormalization-group (RG) invariant pseudoscalar
 currents with quantum numbers of the pion:
\begin{equation}
J_5(x)=\frac{1}{\sqrt{2}}\left(m_u+m_d\right)\left[\bar u(x)i\gamma_5u(x)-
\bar d(x)i\gamma_5d(x)\right]\quad .
\tlabel{current}
\end{equation} 
The particular form of \tref{GenLap} that will be used in obtaining mass bounds is
\begin{equation}
{\cal L}_0^{(5)}(\tau)=\frac{1}{\pi}\int\limits_{4m_\pi^2}^\infty \rho_{_5}(t) \,\mathrm{e}^{-t \tau}\,{\rm d}t \quad .
\tlabel{BasicSr}
\end{equation} 
where $\rho_{_5}(t)$ is the hadronic spectral function appropriate to the pion quantum numbers.

  Perturbative contributions to  ${\cal L}_0^{(5)}(\tau)$ are known up to 
four-loop order in the $\overline{\rm MS}$ scheme \cite{BNRY,chetyrkin}.  
Infinite correlation-length vacuum effects in  ${\cal L}_0^{(5)}(\tau)$
are represented by  the (non-perturbative) QCD condensate contributions \cite{SVZ,BNRY,Bagan}.  
In addition to the QCD condensate contributions, the pseudoscalar (and scalar) correlation functions
are sensitive to finite correlation-length vacuum effects described by direct instantons \cite{dorokhov}
 in the instanton 
liquid model \cite{ins_liquid}.  Combining all these results,  
the total result for ${\cal L}_0^{(5)}(\tau)$ to leading order 
in the light-quark masses is \cite{holder_bounds}
\begin{equation}
\begin{split}
{\cal L}_0^{(5)}(\tau)=&\frac{3m_n^2}{8\pi^2\tau^2}\left(   
1+4.821098 \frac{\alpha}{\pi}+21.97646\left(\frac{\alpha}{\pi}\right)^2+53.14179\left(\frac{\alpha}{\pi}\right)^3
\right)
\\
& +m_n^2\left(
-\langle m\bar q q\rangle 
+\frac{1}{8\pi}\langle \alpha G^2\rangle
+\frac{1}{4}\pi\langle {\cal O}_6\rangle\tau
\right)
\\
 &+m_n^2
\frac{3\rho_c^2}{8 \pi^2\tau^3} \exp{ \left(-\frac{\rho_c^2}{2\tau}\right) }
\left[   
  K_0\left( \frac{\rho_c^2}{2\tau}\right)  +
       K_1\left( \frac{\rho_c^2}{2\tau} \right)
\right]\quad ,
\end{split} 
\tlabel{R5}
\end{equation}
where $\alpha$ and  $m_n=\left(m_u+m_d\right)/2$ are the $\overline{{\rm MS}}$ 
running coupling and quark masses at the scale $1/\sqrt{\tau}$, and $\rho_c=1/(600\,{\rm MeV})$ represents the instanton size in the 
instanton liquid model \cite{ins_liquid}. 
$SU(2)$ symmetry has been used for the dimension-four quark condensates ({\it i.e.} 
$(m_u+m_d)\left\langle \bar u u+ \bar d d\right\rangle\equiv 4 m\left\langle \bar q q\right\rangle$),
and   $\langle {\cal O}_6\rangle$ denotes the dimension six quark condensates 
\begin{equation}
\begin{split}
\langle{\cal O}_6\rangle&\equiv \alpha_s
\biggl[ 
\left(2\langle \bar u \sigma_{\mu\nu}\gamma_5
T^au\bar u \sigma^{\mu\nu}\gamma_5T^a u\rangle
+ u\rightarrow d\right)
 -4\langle \bar u \sigma_{\mu\nu}\gamma_5T^au\bar d 
 \sigma^{\mu\nu}\gamma_5 T^a d\rangle
\biggr. \\
& \qquad\qquad
\biggl.+\frac{2}{3}
\langle \left(   
\bar u \gamma_\mu T^a u+\bar d \gamma_\mu T^a d 
\right)
\sum_{u,d,s}\bar q \gamma^\mu T^aq
\rangle\biggr]\quad .
\end{split}
\tlabel{o6}
\end{equation}
The vacuum saturation hypothesis \cite{SVZ} will be used as a reference value for $\langle{\cal O}_6\rangle$
\begin{equation}
\langle{\cal O}_6\rangle=f_{vs}\frac{448}{27}\alpha
\langle \bar q q\bar q q\rangle
=f_{vs}3\times 10^{-3} {\rm GeV}^6 
\tlabel{o61}
\end{equation}
where $f_{vs}=1$ for exact vacuum saturation.  Larger values 
of effective dimension-six
operators found in \cite{dimsix2,dimsix1} imply that $f_{vs}$  could be as 
large as $f_{vs}=2$.
The quark condensate is determined by the GMOR (PCAC) relation 
\begin{equation}
\left(m_u+m_d\right)\langle \bar u u+\bar d d\rangle=4m\langle \bar q q\rangle=
-2f_\pi^2m_\pi^2
\tlabel{GMOR}
\end{equation}
where $f_\pi=93\,{\rm MeV}$. 

Note that {\em all }the theoretical contributions in \tref{R5} are proportional to  $m_n^2$, demonstrating that the quark mass sets
the scale of the pseudoscalar channel.  This dependence on the quark mass can be singled out as follows:
\begin{equation}
{\cal L}_0^{(5)}(\tau)=\left[m_n\left(1/\sqrt{\tau}\right)\right]^2\,G_5\left(\tau\right)
\tlabel{G5Def}
\end{equation}
where $G_5$ is independent of $m_n$ and is trivially extractable from \tref{R5}.
Higher-loop perturbative contributions in (\ref{R5}) are thus significant since they can 
effectively enhance the quark mass with increasing loop order.  

Determinations of the non-strange quark mass $m_n$ using the sum-rule \tref{BasicSr} require input of a phenomenological model for the spectral function $\rho_{_5}(t)$.  The mass $m_n$ can then be determined by fitting 
to find the best agreement between the phenomenological model and the theoretical prediction 
respectively appearing on the right- and left-hand sides of \tref{BasicSr}.  For example, the simple resonance(s) plus continuum model
\begin{equation}
\frac{1}{\pi}\rho_{_5}(t)=2f_\pi^2m_\pi^4\left[
\delta\left(t-m_\pi^2\right)+\frac{F_\Pi^2 M_\Pi^4}{f_\pi^2m_\pi^4}\delta\left(t-M_\Pi^2\right)
\right]
+\Theta\left(t-s_0\right)\frac{1}{\pi}\rho^{QCD}(t)
\tlabel{ResModel}
\end{equation}
represents the pion pole ($m_\pi$), a narrow-width approximation to the pion excitation ($M_\Pi$) 
such as the $\Pi(1300)$, and a QCD continuum above the continuum threshold $t=s_0$. Of course more 
detailed phenomenological models can be considered which take into account possible width effects for the pion 
excitation, further resonances, resonance(s) enhancement of the $3\pi$ continuum {\it etc.} This leads to significant 
model dependence which partially accounts for the spread of theoretical estimates in \cite{pi_1300,sumrule_mass,3pi}.
Since the common phenomenological portion of all these models is the pion pole, it is valuable to extract quark mass
{\em bounds} which only rely upon the input of the pion pole on the phenomenological side of \tref{BasicSr}.

The existence of such bounds is easily seen by separating the pion pole out from $\rho_{_5}(t)$, in which case 
\tref{BasicSr} becomes
\begin{equation}
\left[m_n\left( 1/\sqrt{\tau} \right) \right]^2 G_5\left(\tau\right)=2f_\pi^2m_\pi^4+
\frac{1}{\pi}\int\limits_{9m_\pi^2}^\infty \rho_{_5}(t) \,\mathrm{e}^{-t\tau}\,{\rm d}t \quad .
\tlabel{PionPoleSr}
\end{equation}
Since $\rho_{_5}(t)\ge 0$ in the integral appearing on the right-hand side of \tref{PionPoleSr}, a bound on the
quark mass is obtained:
\begin{equation}
m_n\left(1/\sqrt{\tau}\right)\ge\sqrt{\frac{2f_\pi^2m_\pi^4}{G_5\left(\tau\right)}}\quad .
\tlabel{SimpleBound}
\end{equation}
Analysis of these bounds following from simple positivity of the ``residual'' portion on the right-hand side of \tref{PionPoleSr} was studied in \cite{bounds,BNRY}.  Other quark mass bounds have been obtained from
dispersion relation inequalities \cite{other_bounds}.

Improvements upon the positivity bound of \tref{SimpleBound} are achieved by developing more stringent inequalities 
based on the positivity of $\rho_{_5}(t)$.
Since $\rho_{_5}(t)\ge 0$, the right-hand (phenomenological) side of 
(\ref{BasicSr}) must satisfy integral inequalities over a measure  ${\rm d}\mu=\rho_{_5}(t)\,{\rm d}t$.
In particular, H\"older's inequality over a measure ${\rm d}\mu$ is \cite{berberian}
\begin{equation}
\biggl|\int_{t_1}^{t_2} f(t)g(t) {\rm d}\mu \biggr|\! \le \!
\left(\int_{t_1}^{t_2} \big|f(t)\big|^ p {\rm d}\mu \right)^{\frac{1}{p}}
\!\!\!\left(\int_{t_1}^{t_2} \big|g(t)\big|^q {\rm d}\mu \right)^{\frac{1}{q}}
~,~
\frac{1}{p}+\frac{1}{q} =1~;~ p,~q\ge 1 \quad ,
\tlabel{HolderIneq}
\end{equation}
which for  $p=q=2$   reduces to the familiar Schwarz inequality, implying 
that the H\"older inequality is a more general constraint. The H\"older inequality can be applied to 
Laplace sum-rules by identifying ${\rm d}\mu=\rho_{_5}(t)\,{\rm d}t$, and defining
\begin{equation}
S_5\left(\tau\right)=\frac{1}{\pi}\int\limits_{\mu_{th}}^\infty \!\!\rho_{_5}(t)\, \mathrm{e}^{-t\tau}\,{\rm d}t
\tlabel{s5}
\end{equation}
where $\mu_{th}$ will later be identified with $9m_\pi^2$.  Suitable choices of 
$f(t)$ and $g(t)$ in the H\"older inequality (\ref{HolderIneq}) yield the following 
inequality  for $S_5(t)$ \cite{sr_holder}: 
\begin{equation}
S_5\left(\tau+(1-\omega)\delta\tau\right)\le \left[S_5\left(\tau\right)\right]^\omega
\left[S_5\left(\tau+\delta\tau\right)\right]^{1-\omega} 
\quad ,~ \forall~ 0\le \omega\le 1\quad .
\tlabel{S5Ineq}
\end{equation}
In practical applications of this inequality,
$\delta\tau\le 0.1\,{\rm GeV^{-2}}$  is used,  in which case this inequality analysis becomes local (depending only on the 
Borel scale $\tau$ and not on $\delta\tau$) \cite{sr_holder,holder_bounds}.

To employ the H\"older inequality (\ref{S5Ineq}) we separate out the pion pole by setting $\mu_{th}=9m_\pi^2$ 
in (\ref{s5})
\begin{equation}
S_5\left(\tau\right)={\cal L}_0^{(5)}(\tau)-2f_\pi^2m_\pi^4=\int\limits_{9m_\pi^2}^\infty \rho_{_5}(t)
 \,\mathrm{e}^{-t\tau}\,{\rm d}t
\tlabel{S5Fin}
\end{equation}
which has a right-hand side  in the standard form \tref{s5} for applying the H\"older inequality. Note that simple positivity of $\rho_{_5}(t)$ gives the inequality
\begin{equation}
S_5\left(\tau\right)\ge 0
\tlabel{positivity}
\end{equation}
which simply rephrases \tref{SimpleBound}.
Lower bounds on the quark mass $m_n$ can now be obtained by finding the minimum value of $m_n$ for which the
H\"older inequality (\ref{S5Ineq}) is satisfied. 
Introducing further phenomenological 
contributions ({\it e.g.} three-pion continuum)  give a slightly larger mass bound as will be discussed later.  
However,  if only the  pion pole is separated out, then the analysis is not subject to uncertainties introduced by the
 phenomenological model.

Although the details are still a matter of dispute, the overall 
validity of QCD predictions at the tau mass is evidenced by the 
analysis of the tau hadronic width, hadronic contributions to $\alpha_{EM}\left(M_Z\right)$ and the 
muon anomalous magnetic moment \cite{braaten_davier}, so 
we impose the inequality (\ref{S5Ineq}) at the tau mass scale  $1/\sqrt{\tau}=M_\tau=1.77\,{\rm GeV}$. 
This also has the advantage of minimizing perturbative uncertainties in the running of $\alpha$ and $m_n$, since the
PDG reference scale for the light-quark masses is at $2\,{\rm GeV}$ \cite{PDG}, in close proximity to $M_\tau$, and 
the result  $\alpha_s\left(M_\tau\right)= 0.33 \pm 0.02$ 
\cite{aleph} can thus be used to its maximum advantage. For the remaining small energy range in which the running of $\alpha$ and $m_n$ is needed, the four-loop $\beta$-function \cite{beta} and four-loop anomalous mass dimension \cite{gamma} 
with three active flavours are used, appropriate to the analysis of \cite{aleph}.  This use of the $2\,{\rm GeV}$ reference scale for $m_n$ 
combined with input of $\alpha\left(M_\tau\right)$ improves upon the perturbative uncertainties in 
\cite{holder_bounds} which employed $\alpha\left(M_Z\right)$ and a $1\,{\rm GeV}$ $m_n$ reference scale which 
necessitated matching through the (uncertain) $b$ and $c$ flavour thresholds.

Further theoretical uncertainties devolve from the QCD condensates
as given in \tref{aGG} and \tref{o61} with $1\le f_{vs}\le 2$, along with a 15\% uncertainty in the
instanton liquid parameter $\rho_c$ \cite{ins_liquid}. The effect of higher-loop perturbative contributions to ${\cal L}_0^{(5)}(\tau)$ on the
resulting $m_n$ bounds is estimated using an asymptotically-improved Pad\'e estimate \cite{apap} of the five-loop term, introducing a $138\left(\alpha/\pi\right)^4$ correction into \tref{R5}. Finally, we allow for the 
possibility that the overall scale of the instanton is 50\% uncertain.

The resulting H\"older inequality bound on the  $2.0\,{\rm GeV}$ $\overline{\rm MS}$ quark masses, updating the
analysis of \cite{holder_bounds},  is
\begin{equation}
m_n(2\,{\rm GeV})=\frac{1}{2}\left[ m_u(2\,{\rm GeV})+m_d(2\,{\rm GeV})\right]\ge 2.1\,{\rm MeV}
\quad .
\tlabel{FinalPDGBound}
\end{equation}
This final result is identical to previous bounds on  $m_n(1\,{\rm GeV})$ \cite{holder_bounds} after conversion to 
$2\,{\rm GeV}$ by the PDG \cite{PDG}, indicative of the consistency of perturbative inputs used in the two analyses.
The theoretical uncertainties in the quark mass bound (\ref{FinalPDGBound}) from the QCD parameters and 
(estimated) higher-order perturbative effects are less than 10\%, and the result (\ref{FinalPDGBound}) 
is the absolute lowest bound resulting from  the  uncertainty analysis.  The dominant sources of uncertainty are $\alpha\left(M_\tau\right)$ and potential higher-loop corrections.  The instanton size $\rho_c$ is the major source of non-perturbative uncertainty, but its effect is smaller than the perturbative sources of uncertainty.

Compared with the
positivity inequality  \tref{positivity}, as first used to obtain quark mass bounds from QCD 
sum-rules \cite{bounds,BNRY},
the H\"older inequality leads to  quark mass bounds 50\% larger   for
identical theoretical and phenomenological inputs at $1/\sqrt{\tau}=M_\tau$, demonstrating that the H\"older inequality 
provides  stringent constraints on the quark mass.  

Finally, we discuss the effects of extending the resonance model to include the $3\pi$ continuum calculated using lowest-order chiral perturbation theory \cite{3pi}  
\begin{gather}
\frac{1}{\pi}\rho_{_5}(t)=2f_\pi^2m_\pi^4\left[
\delta\left(t-m_\pi^2\right)+\Theta\left(t-9m_\pi^2\right)\rho_{3\pi}(t)\frac{t}{18\left(16\pi^2f_\pi^2\right)^2}\right]
\\
\begin{split}
\rho_{3\pi}(t)=&\int\limits_{4m_\pi^2}^{\left(\sqrt{t}-m_\pi\right)^2}
\frac{{\rm d}u}{t}\sqrt{\lambda\left(1,\frac{u}{t},\frac{m_\pi^2}{t}\right)}\sqrt{1-\frac{4m_\pi^2}{u}}
\Biggl\{
5+\Biggr.
\\
 &+\frac{1}{2\left(t-m_\pi^2\right)^2}\left[\frac{4}{3}\left(t-3\left(u-m_\pi^2\right)\right)^2
+\frac{8}{3}\lambda\left(t,u,m_\pi^2\right)\left(1-\frac{4m_\pi^2}{u}\right)+10m_\pi^4
\right]
\\
 &\Biggl.+\frac{1}{t-m_\pi^2}\left[3\left(u-m_\pi^2\right)-t+10m_\pi^2
\right]\Biggr\}
\end{split}
\tlabel{3piCont}
\\
\lambda(x,y,z)=x^2+y^2+z^2-2xy-2yz-2xz
\end{gather}
which becomes $\rho_{3\pi}(t)\rightarrow 3$ in the limit $m_\pi\rightarrow 0$ .  
Inclusion of the $3\pi$ continuum \tref{3piCont} 
is still likely to underestimate the total spectral function since more 
complicated models of the spectral function involve resonance enhancement of this $3\pi$ continuum \cite{3pi}.
If this limiting form is used up to a cutoff
of $1\,{\rm GeV}$, then the resulting H\"older inequality quark mass bounds are {\em raised} by approximately 10\%, and  
a 14\% effect is observed if the cutoff is moved to infinity.\footnote{The exponential suppression of the large-$t$ region
in the Laplace sum-rule
\protect\tref{BasicSr} minimizes any errors in this region from this approximation to the  $3\pi$ continuum, and also leads to the observed small difference in extending the cutoff to infinity.}  
Working with the full form \tref{3piCont} complicates the numerical analysis, but the  following simple form (with $t$ in GeV units) is easily verified to be a bound on the $3\pi$ continuum in the region below $1\,{\rm GeV}$:
\begin{equation}
\rho_{3\pi}(t)\ge\frac{4}{3}\left[\left(\sqrt{t}-m_\pi\right)^2-4m_\pi^2 \right]\quad .
\tlabel{rhoAp}
\end{equation}
This approximate form of the $3\pi$ continuum again raises the resulting quark mass bounds by approximately 10\%.

\vskip 1cm
{\noindent \large\bf Acknowledgments:}\\
Many thanks to Professor Sugamoto and his group for organizing a valuable and enjoyable workshop.
This research was supported by   the Natural Sciences and Engineering Research Council of Canada (NSERC).

\end{document}